\documentclass{aa}  

\usepackage{graphicx}
\usepackage{realboxes}
\usepackage{txfonts}
\usepackage{amssymb}
\usepackage{natbib}
\usepackage{hyperref}
\usepackage{pdflscape}
\usepackage{hyperref}
\usepackage[normalem]{ulem}
\usepackage{xcolor}
\begin{document}

\title{Normalized angular momentum deficit: A tool for comparing the violence of the dynamical histories of planetary systems}

\titlerunning{Normalized angular momentum deficit}
   
   \author{D. Turrini
          \inst{1}
          \and
          A. Zinzi\inst{2}
          \and
          J.~A. Belinchon\inst{3}                 
          }

   \institute{Institute for Space Astrophysics and Planetology INAF-IAPS, Via Fosso del Cavaliere 100, 00133, Rome, Italy\\
              \email{diego.turrini@inaf.it}
              \and
              Space Science Data Center (SSDC) - ASI, Via del Politecnico snc, 00133, Rome, Italy\\
              \email{angelo.zinzi@ssdc.asi.it}
              \and
              Departamento de Matem\'{a}ticas, Universidad de Atacama, 485 Copayapu, Copiap\'o, Chile\\
              \email{jose.belinchon@uda.cl}
             }

   \date{Received XXXX; accepted XXXX}

% \abstract{}{}{}{}{} 
% 5 {} token are mandatoryAIMS
 
  \abstract
   {Population studies of the orbital characteristics of exoplanets in multi-planet systems have highlighted the existence of an anticorrelation between the average orbital eccentricity of planets and the number of planets of their host system, that is, its multiplicity. This effect was proposed to reflect the varying levels of violence in the dynamical evolution of planetary systems.}
   {Previous work suggested that the relative violence of the dynamical evolution of planetary systems with similar orbital architectures can be compared through the computation of their angular momentum deficit (AMD). We investigated the possibility of using a more general metric to perform analogous comparisons between planetary systems with different orbital architectures.}
   {We considered a modified version of the AMD, the normalized angular momentum deficit (NAMD), and used it to study a sample of 99 multi-planet systems containing both the currently best-characterized extrasolar systems and the solar system, that is, planetary systems with both compact and wide orbital architectures.}
   {We verified that the NAMD allows us to compare the violence of the dynamical histories of multi-planet systems with different orbital architectures. We identified an anticorrelation between the NAMD and the multiplicity of the planetary systems, of which the previously observed eccentricity--multiplicity anticorrelation is a reflection.}
   {Our results seem to indicate that phases of dynamical instabilities and chaotic evolution are not uncommon among planetary systems. They also suggest that the efficiency of the planetary formation process in producing high-multiplicity systems is likely to be higher than that suggested by their currently known population.}


   \maketitle
%
%________________________________________________________________

\section{Introduction}\label{section-introduction}

The ever-growing catalog of exoplanets discovered to date contains about 4000 planets distributed in about 3000 planetary systems. At least one in five of these planetary systems is confirmed to contain between 2 and 7 planets, i.e., to have multiplicity $M > 1$, although their orbital and physical characteristics are still unevenly known and are often affected by large uncertainties. Notwithstanding these limitations, the known population of multi-planet systems already showcases the diversity of architectures that can arise from the planetary formation process and the dynamical evolution of planetary systems. 

The available orbital and physical data, while incomplete, enabled the first population studies on statistically significant samples of planets. The analysis of the distribution of orbital eccentricities of exoplanets as a function of the multiplicity of their host systems, in particular, revealed the existence of an anticorrelation between the average orbital eccentricity $\overline{e}$ and the planetary multiplicity \citep{juric2008,limbach2015,zinzi2017}. This means that, on average, the orbits of planets in systems with lower multiplicity (e.g., $M=2$ or $3$) are characterized by higher eccentricities than those of planets in systems with higher multiplicity (e.g., $M\ge4$).

\begin{figure}[t]
\centering
\includegraphics[width=\columnwidth]{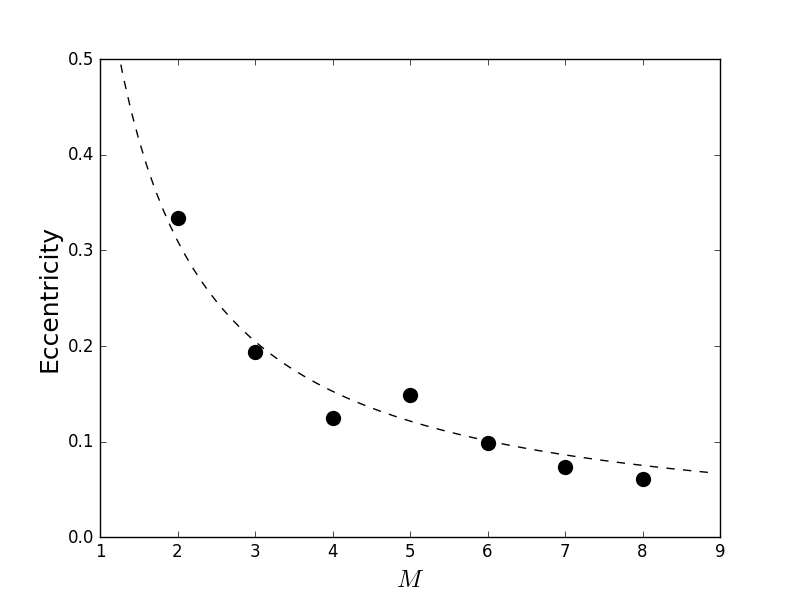}
\caption{Average eccentricity $\overline{e}$ vs. multiplicity $M$ anticorrelation reported by \citet{zinzi2017} studying the sample of the 258 more completely characterized exoplanets at the time. The symbols for $M=7$ and $M=8$ are TRAPPIST-1 and solar system, respectively. The current average eccentricity of TRAPPIST-1 is lower than the value estimated from the data available in 2017 (see also Sect. \ref{section-method}).}
\label{Figure1}%
\end{figure}

While early results suggested this anticorrelation to break for systems with low multiplicity ($ M \leq 3$, \citealt{limbach2015}), more recent analyses accounting for the uncertainties on the estimated eccentricities revealed (see Fig. \ref{Figure1}) a trend extending from $M=2$ to $M=8$, i.e., the solar system \citep{zinzi2017}. To find a physical interpretation to this anticorrelation, \citet{zinzi2017} investigated the subset of multi-planet systems for which all the physical parameters needed to compute their angular momentum were known; that is, stellar mass and planetary masses, semimajor axes, eccentricities, and inclinations. 

For those systems, these authors computed the angular momentum deficit (AMD), a quantity devised to study the dynamical stability of the solar system \citep{laskar1997,laskar2000} and more recently applied to the analysis of the dynamical stability of exoplanetary systems \citep{laskar2017}. Even with the limited number of multi-planet systems that could be included in their analysis, \citet{zinzi2017} observed an analogous anticorrelation between the AMD and the multiplicity.

The AMD is defined as the difference between the angular momentum of an idealized system, with the same planets of the real system orbiting at the same semimajor axes from the star on circular and planar orbits, and the norm of the angular momentum of the real planetary system (i.e., the projection of its angular momentum on the invariable plane; see Sect. \ref{section-method} and \citealt{laskar1997,laskar2000,laskar2017} for details).

The invariable planes of the known exoplanetary systems are unknown and the inclination values of their planets are measured with respect to the line of sight. To circumvent these limitations, \cite{zinzi2017} adopted  the orbital plane of the most massive planet in each planetary system as the local reference plane. The reasoning behind this choice is that, due to the larger inertial mass of the planet, its orbit is likely the least perturbed in the system \citep[see also][]{laskar2017}). In computing the AMD of the system, all the planetary inclinations were changed into relative inclinations with respect to this reference plane.

The effect of non-zero eccentricities and inclinations is that of lowering the norm of the orbital angular momentum \citep{laskar1997,laskar2017}. Therefore, the AMD is a positive-defined quantity that increases the more the orbits deviate from the circular and planar case. The positive-defined nature of the AMD also means that the presence of undiscovered planets in a planetary system can either increase the AMD or leave it unaffected. The latter case occurs when the undiscovered planet is on a circular and planet orbit, so that its contribution to the AMD of the system is zero. 

The works of \citet{chambers2001} and \citet{laskar2017} showed how chaotic diffusion and dynamical friction between planetary bodies lead to an increase of their dynamical excitation and, consequently, their AMD particularly during the early stages of the formation and evolution of a planetary system. Mutual collisions between excited planetary bodies \citep{chambers2001,laskar2017} and their removal from the system by ejections (e.g., by planet-planet scattering events; \citealt{weidenschilling1996,rasio1996,marzari2002,chatterjee2008}) and collisions with the host star \citep{chambers2001} act instead to reduce the AMD and stabilize the planetary system.

As a result of these stabilizing processes, after its initial growth following the onset of a phase of dynamical excitation, the AMD of a planetary system decreases to a smaller but non-zero value (see, e.g., Fig. 6 in \citealt{chambers2001}). Planetary systems that have experienced or are experiencing phases of violent and chaotic dynamical evolution can thus be expected to have higher AMD values than stable systems that did not undergo similar phases, even after the loss (by ejection or collision with the host star; \citealt{chambers2001}) of their most dynamically excited bodies. 

As a consequence, \citet{zinzi2017} argued that the comparison of the AMD values of planetary systems could offer an indication of the relative violence of their dynamical histories. These authors also argued that, if confirmed, the anticorrelation between AMD and multiplicity could indicate that the currently known population of high-multiplicity systems on average experienced less violent dynamical histories than the population of low-multiplicity systems \citep{zinzi2017}. 

This interpretation, however, is valid only if the sample of planetary systems under study is characterized by similar orbital architectures, as was the case of the restricted sample considered by \citet{zinzi2017}. The AMD values of planetary systems with significantly different orbital structures and planetary masses are not easily comparable. All other parameters being equal, planetary systems with planets on wide orbits and/or massive planets easily have larger AMD values than more compact planetary systems with less massive planets even if the eccentricities and inclinations of the latter are higher than those of the former. 

In the specific case of the solar system, in particular, even if the orbits of its planets are almost circular and planar its AMD is higher than those of the exoplanetary systems in the sample considered by \citet{zinzi2017} because of their compact architectures. Sect. \ref{section-method} provides further discussion on the specificity of the solar system in the context of our work.

The capability of comparing the relative violence of the dynamical history of planetary systems would provide important insights into how planetary systems form and evolve and would prove critical to link the composition of planets to their formation history (e.g., \citealt{madhusudhan2016,turrini2018}), particularly in view of the future observations by the James Webb Space Telescope (e.g., \citealt{cowan2015}) and the space mission ARIEL (Atmospheric Remote-sensing Infrared Exoplanet Large-survey, \citealt{tinetti2018,turrini2018}) of the European Space Agency.

This work investigates a modified version of the AMD, the normalized angular momentum deficit (NAMD), to assess whether it can be used to analyze and compare populations of planetary systems characterized by very different architectures. The rest of the work is organized as follows. In Sect. \ref{section-method} we describe the computation of the NAMD and its uncertainty. In Sect. \ref{section-results} we describe the sample of planetary systems we considered and the distribution of their NAMD values. In Sect. \ref{section-conclusions} we discuss our results and draw the conclusions of this work.

%________________________________________________________________

\section{Methods}\label{section-method}

As defined by \citet{laskar1997,laskar2000} and \citet{laskar2017}, the  $AMD_{k}$ of a individual planetary body can be expressed as
\begin{equation}
AMD_{k} = m_{k} \sqrt{G m_{0}  a_{k}}  \left(1-\sqrt{1-e_{k}^{2}}  \cos\,i_{k}\right)
\label{eqn-AMD_single}
,\end{equation}
where the independent variables are the planetary mass $m_{k}$, semimajor axis $a_k$, orbital eccentricity $e_k$, orbital inclination $i_k$, and stellar mass $m_{0}$;  $G$ is the gravitational constant.

The total  $AMD$ of a planetary system with multiplicity $M$ is given by the sum of Eq. \ref{eqn-AMD_single} over all its planets, i.e., by the sum of $AMD_k$ over all $k$ indexes \citep{laskar1997,laskar2000,laskar2017} as follows:

\begin{eqnarray}
AMD & = & \sum_{k=1}^{M} AMD_{k} = \nonumber \\
& = & \sum_{k} m_{k} \sqrt{G  m_{0}  a_{k}}  \left(1-\sqrt{1-e_{k}^{2}}  \cos\,i_{k}\right)
\label{eqn-AMD}
.\end{eqnarray}

From Eqs. \ref{eqn-AMD_single} and \ref{eqn-AMD}, it is immediately evident  that the AMD of planetary systems with significantly different architectures (e.g., wide versus compact and massive planets versus small planets) cannot be directly compared. The scaling on the semimajor axis and the planetary mass changes the absolute value of the AMD and could mask the effects of the orbital eccentricities and inclinations. The use of the AMD as a comparative measure between various planetary systems, consequently, is meaningful only if samples homogeneous in terms of architecture are considered. 

To overcome this limitation we focused on the relative deficit of angular momentum instead of the absolute one by considering the NAMD. The NAMD is a quantity introduced by \citet{chambers2001} to compare the outcomes of simulations of the formation of terrestrial planets in the solar system and is defined as the ratio of the AMD to the angular momentum of the idealized planetary system with the same masses and semimajor axis values for the planets, but with circular and co-planar orbits (hereafter referred to as CAM, circular angular momentum), that is,

\begin{eqnarray}
NAMD & = & \frac{AMD}{CAM} = \frac{AMD}{\sum_{k} m_{k} \sqrt{G  m_{0}  a_{k}}} = \nonumber \\
& = & \frac{\sum_{k} m_{k} \sqrt{a_{k}}  \left(1-\sqrt{1-e_{k}^{2}}  \cos\,i_{k}\right)}{\sum_{k} m_{k} \sqrt{a_{k}}}
,\end{eqnarray}

where the dependence from the stellar mass $m_0$ and the gravitational constant $G$ are removed \citep{chambers2001}.

To build our sample of multi-planet systems we queried the NASA Exoplanet Archive\footnote{\url{https://exoplanetarchive.ipac.caltech.edu}} \citep{akeson2013}  
through the ExoplAn3T online tool\footnote{\url{https://tools.ssdc.asi.it/exoplanet}} (Exoplanet Analysis and 3D Visualization Tool) of the Space Science Data Center (SSDC) of the Italian Space Agency (ASI). 

In our query we gathered the planetary systems for which masses, semimajor axes, and eccentricities were available for all planets. When the inclination values were available for all planets, we followed \citet{zinzi2017} and computed the relative inclinations with respect to the most massive planet in the system. Barring individual cases with specific viewing geometries (e.g., if the line of nodes is along the line of sight), this approach provides a reasonable overall estimate of the mutual inclinations in our population of systems.

When the inclinations were not available or incomplete we assumed the planets to be on coplanar orbits to increase the otherwise limited statistics; see below for the treatment of the uncertainties in this case. This approach allowed us both to extend the sample of planetary systems in our analysis and to minimize the impact of systems with unfavorable viewing geometries.

The data collected for the various exoplanetary systems are updated to November 19, 2019, with the exception of TRAPPIST-1 for which we adopted the values of the planetary masses, semimajor axes, and eccentricities estimated by \citet{grimm2018}, as they are more complete and precise than those currently present in the NASA Exoplanet Archive.%}

As a consequence of the incompleteness and inhomogeneity of exoplanetary data, we need to account for the different uncertainties with which the planetary parameters are known to properly compare the NAMD of different planetary systems. We also need to test whether the anticorrelation between multiplicity and AMD found by \cite{zinzi2017} is associated with a real effect, that is, not an artifact of the architectures of the specific sample considered, and if this holds when systems with heterogenous architectures are included in the sample. 

To estimate the uncertainty on the NAMD values of the planetary systems we followed the Monte Carlo approach detailed by \citet{laskar2017} and performed 10000 Monte Carlo extractions for each planetary parameter. For the masses and semimajor axes we assumed Gaussian distributions centered on the values reported in the databases and standard deviations equal to half the reported confidence interval. All these distributions were truncated to zero \citep{laskar2017}.

For those systems for which we possess inclination values for all planets, we assumed Gaussian distributions centered on the relative inclinations with respect to the most massive planet in the system and standard deviations equal to half the reported confidence interval.

For those systems for which we do not possess inclination values for all planets, we assumed Gaussian distributions centered on zero and standard deviations equal to the modulus of the arcsin of the eccentricity $\sigma_{i} = \left| \arcsin e \right|$. The latter relationship arises from assuming the equipartition of the dynamical excitation $\sqrt{e^2 + \sin^2 \left( i \right)}$ \citep{petit2001} along all secular degrees of freedom of the system \citep{laskar2017}.

Finally, for the eccentricities we once again followed the approach described by \citet{laskar2017}. We focused on the rectangular eccentricity coordinates $\left( e\cos\omega, e\sin\omega  \right)$ and assumed Gaussian distributions for both. As the average value of $\omega$ does not influence the computation of the eccentricity distribution, its value was set to $\omega=0$ \citep{laskar2017}.

As a result, the distribution of $\tilde{e}=e\cos\omega$ is centered on the mean value $\langle e \rangle$ of the eccentricity reported in the catalogs, while the distribution of $e\sin\omega$ is centered on zero. The standard deviation of both distributions was set equal to half the reported eccentricity confidence interval. The distribution of $e$ can then be derived from that of $\tilde{e}$ through the relationship $e=\sqrt{\tilde{e}^2 + \left( \tilde{e} - \langle e \rangle \right)^2}$.

The values and relative uncertainty of the NAMD of each planetary system can then be used to compute the weighted average NAMD of the planetary systems falling in each multiplicity bin, similarly to what was done by \cite{zinzi2017} in computing the average eccentricity of each multiplicity bin (see Eqs. 1 and 2 in \citealt{zinzi2017}). 

We did not consider in our analysis any systems for which the relative uncertainties on the NAMD value resulted of $1$ or larger, that is, the uncertainty is of the same order as, or larger than, the measured value. We note that, to limit the effects of small numbers statistics, we combined the cases $M=4$ and $M=5$ into a single population associated with a multiplicity $M=4.5$. Similarly, we combined the cases $M=6$ and $M=7$ into a unique population associated with a multiplicity $M=6.5$.

To test the effectivess of the NAMD as a comparative measure, we also included the solar system ($M=8$) in our sample using the values of its parameters reported by NASA\footnote{\url{https://nssdc.gsfc.nasa.gov/planetary/factsheet/}}. Given that the uncertainties on the orbital and physical parameters of the planets in the solar system are much smaller than those of the exoplanetary systems we considered, we treated the values of these parameters as exact and did not propagate their uncertainties.

Before presenting the results of our analysis, the cases of the solar system and TRAPPIST-1 require further discussion. Dynamical studies of its long-term stability have shown how the solar system can be considered as the sum of two separate subsystems: the inner, AMD-unstable system encompassing the terrestrial planets and the outer, AMD-stable system encompassing the giant planets \citep{laskar1997,laskar2000,laskar2017}. Furthermore, the solar system has been suggested to have undergone past episodes of dynamical instability \citep[e.g.,][and references therein]{nesvorny2018}.

Dynamical studies of the formation of the compact system around TRAPPIST-1 highlighted instead the role of orbital resonances and tidal forces in governing its long-term evolution and stability \citep[e.g.,][and references therein]{tamayo2017,papaloizou2018}. We therefore know that the solar system and TRAPPIST-1 had complex dynamical histories characterized by evolutionary tracks of opposite nature: the first was marked by the presence of chaos while the second was shaped by order.

In our analysis we ignored this a priori information and simply include these two systems as members of our population, as a further test of the usefulness of the NAMD as a comparative measure of the violence of the dynamical history. As we also discuss in Sect. \ref{section-conclusions}, however, the coupling in future works of the NAMD analysis with the AMD-stability study \citep{laskar2017} allows for a more complete characterization of the dynamical state of the systems being compared and a more detailed understanding of the root of their differences.
 
%________________________________________________________________

\section{Results}\label{section-results}

\begin{figure}
\centering
\includegraphics[width=\columnwidth]{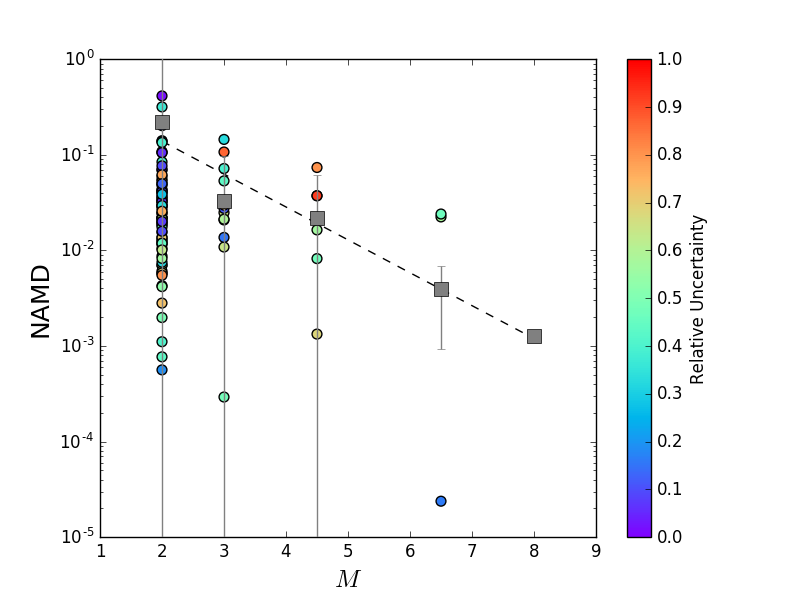}
\caption{Multiplicity vs. NAMD for the individual planetary systems considered in this study (colored circles without error bars) and for the average values of each multiplicity bin (gray squares with error bars). The color code for the individual planetary systems indicates the relative uncertainty of their NAMD value. The linear fit is shown only as a visual aid.}
\label{Figure2}%
\end{figure}

The sample of multi-planet exoplanetary systems resulting from our query and possessing relative uncertainties on the NAMD lower than 1 is composed of $98$ planetary systems. The sample contains a total of $234$ exoplanets, plus the solar system, as follows:
\begin{itemize}
\item 77 systems with two planets ($M=2$);
\item 12 system with three planets ($M=3$); 
\item 5 systems with four planets ($M=4$);
\item 1 system with five planets ($M=5$);
\item 2 systems with six planets ($M=6$);
\item 1 system with seven planets ($M=7$, TRAPPIST-1);
\item 1 system with eight planets ($M=8$, the solar system);
\end{itemize}
where, as discussed at the end of Sect. \ref{section-method}, we combined the cases $M=4$ and $M=5$ into a single bin of 6 systems with $M=4.5$ and the cases $M=6$ and $M=7$ into a single bin of 3 systems with $M=6.5$ to limit the effects of small numbers statistics.

The results of our analysis are shown in Fig. \ref{Figure2}. As can be seen, the individual systems in each multiplicity bin show a spread in their NAMD values as a result of their different dynamical histories. TRAPPIST-1 has the lowest NAMD value both in the multiplicity bin $M=6.5$ and in the whole sample, possibly as a result of the tidal circularization of the compact orbits of its planets \citep{papaloizou2018}.

Nevertheless, the weighted-average NAMD values we computed for each bin show a decreasing trend with increasing multiplicity values, i.e., an anticorrelation between NAMD and multiplicity. This decreasing trend extends down to $M=8$, that is, the solar system in constrast to the case of the AMD \citep{zinzi2017}. In Fig. \ref{Figure2} we also show a root-mean-squared fit of the average NAMD values in our sample; however, this fit is intended only as a visual aid because of the still limited statistics of systems with higher multiplicity values ($M\geq 4$).

\begin{figure}
\centering
\includegraphics[width=\columnwidth]{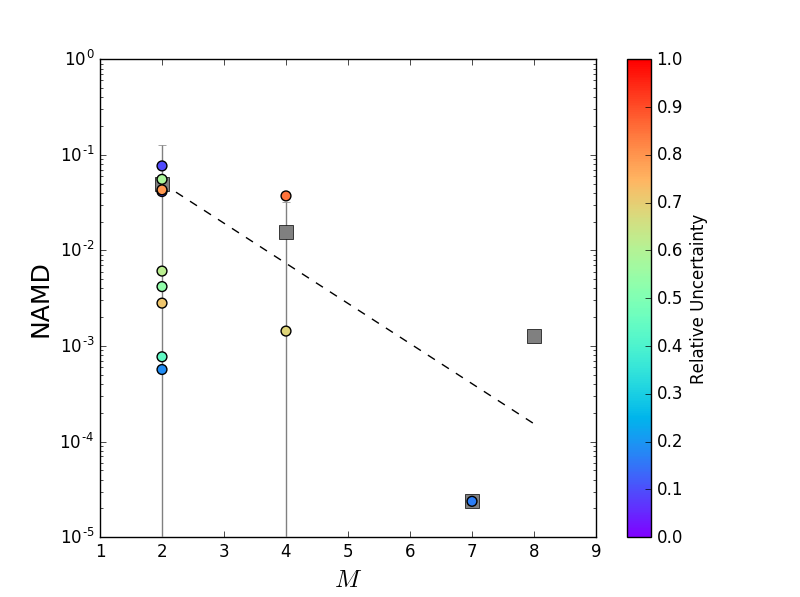}
\caption{Multiplicity vs. NAMD\ for the individual planetary systems (colored circles without error bars) and for the average values of each multiplicity bin (gray squares with error bars) of the subsample of systems for which all planetary parameters, including inclination, are known. The color code for the individual planetary systems indicates the relative uncertainty of their NAMD value. The linear fit is shown only as a visual aid.}
\label{Figure3}%
\end{figure}

To test for possible artifacts generated by our treatment of the unknown planetary inclinations, we performed the same analysis on the subsample of systems for which also the values of the inclination are known (see Sect. \ref{section-method}). This subsample consists of 12 exoplanetary systems, containing a total of 33 exoplanets, plus the solar system, as follows:
\begin{itemize}
\item 9 systems with two planets ($M=2$);
\item 2 systems with four planets ($M=4$);
\item 1 system with seven planets ($M=7$, TRAPPIST-1);
\item 1 system with eight planets ($M=8$, the solar system); 
\end{itemize}The results of our analysis are shown in Figure \ref{Figure3}, while the names of the systems in this subsample, together with their multiplicity, their NAMD values, and the associated relative uncertainties are reported in Table \ref{Table1}. The NAMD-$M$ anticorrelation appears to be broadly preserved in this subsample, even if the systems it contains are limited in number and sparse.  

\begin{table}[t]
\centering
\caption{Multiplicity, NAMD values, and associated relative uncertainties for the subsample of multi-planet extrasolar systems for which all planetary parameters, including inclination, are known.}
\label{Table1}
\begin{tabular}{|c|c|c|c|}
\hline
Planetary & $M$ & NAMD & $\sigma_{rel}$ \\
System    &     &      &                \\ 
\hline
HD 15337  &  2  &  $6.08\times10^{-3}$  &  0.62  \\
Kepler-117  &  2  &  $5.68\times10^{-4}$  &  0.19  \\
Kepler-87  &  2  &  $7.72\times10^{-4}$  &  0.44  \\
7 CMa  &  2  &  $4.19\times10^{-3}$  &  0.53  \\
HD 202696  &  2  &  $2.81\times10^{-3}$  &  0.72  \\
HD 192310  &  2  &  $5.55\times10^{-2}$  &  0.58  \\
HD 82943  &  2  &  $4.13\times10^{-2}$  &  0.09  \\
HD 106315  &  2  &  $4.30\times10^{-2}$  &  0.79  \\
Kepler-419  &  2  &  $7.66\times10^{-2}$  &  0.10  \\
Kepler-79  &  4  &  $1.43\times10^{-3}$  &  0.68  \\
KOI-94  &  4  &  $3.72\times10^{-2}$  &  0.85  \\
TRAPPIST-1  &  7  &  $2.38\times10^{-5}$  &  0.17 \\
Solar system  &  8  &  $1.27\times10^{-3}$  &  - \\
\hline
\end{tabular}
\end{table}

\section{Conclusions}\label{section-conclusions}

In this work we investigated the possibility of comparing the relative violence of the dynamical histories of planetary systems through the information on the evolution of their angular momentum recorded in their present orbital architectures. 

Previous work focusing on planetary systems with similarly compact orbital architectures suggest that such a comparison is in principle possible by considering the differences in their AMD values \citep{zinzi2017}. However, in case of very different architectures (e.g., Kepler-419 versus the solar system, see Table \ref{Table1}) the comparison of their AMD values does not provide us with meaningful information (see Sects. \ref{section-introduction} and \ref{section-method} and \citealt{zinzi2017}).

We therefore focused on the metric provided by the normalized version of the AMD, the NAMD, previously introduced by \citet{chambers2001}. We used the NAMD to study a sample of planetary systems with heterogeneous orbital architectures containing both the solar system and the 98 multi-planet exosystems more completely characterized to date. Our results confirm the possibility of using the NAMD to systematically compare planetary systems with very different orbital architectures.

Before discussing the results of our analysis beyond the use of the NAMD as a comparative metric, we want to point out that the statistics of high--multiplicity systems is still limited and the available systems are sparsely distributed over a range (4 $\leq$ M $\leq$ 8) of multiplicity values. The combination of observational biases and still limited statistics might result in the under-representation of high-M, high-NAMD systems in the currently known population, as larger eccentricities should be associated with larger orbital separations for the systems to be stable, thus lowering the detectability of these systems.

Furthermore, as discussed by \citet{zhu2019} the current characterization of the architecture of the multi-planet systems discovered by the NASA mission Kepler could be the result of detection biases, at least in terms of the planetary masses and orbital spacing. Given that the sample population used in our study has been characterized both from space and from ground using multiple techniques, however, we expect the latter effect to have a limited impact on our results.

Bearing these caveats in mind, our analysis reveals an anticorrelation between NAMD and multiplicity $M$ extending from exosystems with two planets ($M=2$) to the solar system ($M=8$), confirming the similar trend observed by \citet{zinzi2017} when studying exoplanetary systems characterized by compact architectures. This anticorrelation suggests that the known high-multiplicity systems experienced, as a whole, less violent dynamical histories than their low-multiplicity counterparts.

The NAMD--$M$ anticorrelation appears both when considering the full sample of 98 exoplanetary systems for which planetary masses, semimajor axes, and eccentricities are known and the more sparse subsample of 12 exoplanetary systems for which the inclinations are also known (i.e., all parameters needed to compute the NAMD without additional assumptions), confirming that the anticorrelation is not an artifact of our assumptions on the unknown inclinations in the larger sample.

The NAMD offers a comparative measure of the dynamical excitation of planetary systems, which in turn is linked to the violence of their past dynamical evolution (see also \citealt{chambers2001} and \citealt{laskar2017}). As an example, the comparison of the NAMD values of the solar system and TRAPPIST-1 suggests that the first was characterized by a relatively more violent or chaotic dynamical history than the second. This is consistent with the results of dynamical studies of the AMD-stability of the solar system \citep{laskar1997,laskar2000,laskar2017} and the formation and evolution of TRAPPIST-1 \citep{tamayo2017,papaloizou2018}.

Notwithstanding this, care should be taken in not overinterpreting the information content of the NAMD. As an illustrative example, the NAMD does not track changes in the semimajor axes due to planet-planet scattering, particularly if they are associated with orbital circularization from dynamical friction or tidal effects. Furthermore, the NAMD does not provide direct evidence of the previous loss (by collision or by ejection) of planetary bodies from the system.

As a result, comparing the NAMD of planetary systems does not allow for discriminating previously quiescent systems that recently entered phases of violent, chaotic dynamical evolution from systems that already underwent similar phases and got stabilized by the loss of some of their original planets. This limitation, however, can be overcome by coupling the NAMD analysis of a sample of planetary systems to that of their AMD-stability \citep{laskar2017}.

It is interesting to note that by construction the NAMD is a function of the eccentricity of the planets populating it (weighted over their masses and semimajor axes). This means that the observed anticorrelation $\overline{e}$-$M$ between the average eccentricity of planets in multi-planet systems and the multiplicity $M$ of the latter, which was observed studying samples of a few hundreds of exoplanets (\citealt{juric2008,limbach2015,zinzi2017}, see Fig. \ref{Figure1}), is simply a reflection of the NAMD--$M$ anticorrelation.

From a more general point of view, our results support the idea that phases of dynamical instabilities and chaotic evolution are not uncommon among planetary systems, as also suggested by the results of the AMD--stability analysis of \citet{laskar2017}. The fact that less populated systems ($M\leq3$) possess, on average, higher NAMD values than more populated systems ($M\geq4$) reveals that the former underwent more violent dynamical histories than the latter. The large fraction of low multiplicity systems with high NAMD values (e.g., above a few $10^{-2}$, which means higher than the average NAMD of systems with $M\geq4$), in turn, suggests that this kind of violent evolution track is reasonably frequent among planetary systems.

The joint effect of a widespread occurrence of phases of violent chaotic dynamical evolution with the stabilizing effects of mutual collisions and the removal of planetary bodies \citep{laskar1997,chambers2001,laskar2017} could imply that the currently observed population of planetary systems in each given multiplicity bin is a combination of two different populations. An original population of systems formed with that specific multiplicity and a population of systems formed with higher multiplicity that lost some of their planets, either collisionally or by ejection, during their evolution.

Should this be true, the planetary formation process would appear to favor the creation of systems with higher multiplicity than revealed by the currently known sample. It is interesting to note that scenarios invoking the loss of additional giant planets have also been formulated in recent years for the solar system (e.g., \citealt{nesvorny2018} and references therein).

Finally, once considered in the framework of recent dynamical stability studies of known exoplanets (\citealt{agnew2019} and references therein), our results open up the possible use of the NAMD as a way to identify promising systems for follow-up observations aimed at finding undiscovered planetary companions. These studies show how even moderate eccentricities can have disruptive effects on multi-planet systems so, within a given sample of potential targets, systems characterized by higher NAMD values would prove less favorable environments for the presence of additional planets, while those characterized by lower NAMD values could represent the best candidates for future searches of new planets.

\begin{acknowledgements}

The authors wish to thank an anonymous referee for comments and suggestions that helped to improve the quality and results of this work. D.T. acknowledges the support of the Italian Space Agency (ASI) through the ASI-INAF contract 2018-22-HH.0 and the Italian National Institute of Astrophysics (INAF) through the INAF Main Stream project ``ARIEL and the astrochemical link between circumstellar discs and planets''. A.Z. acknowledges financial support from the ASI-INAF agreement 2018-16-HH.0. J.A.B. was supported by the COMISION NACIONAL DE CIENCIAS Y TECNOLOGIA through FONDECYT Grant N0 11170083. This research has made use of the NASA Astrophysics Data System Bibliographic Services.
\end{acknowledgements}

%-------------------------------------------------------------------

\end{document}